\documentclass{ws-procs10x7} 
\usepackage{balance} 
\usepackage{color,psfig} 
 
\newcolumntype{d}[1]{D{.}{.}{#1}} 
 
\makeindex 
\begin{document} 
 
\title{Lattice QCD with light twisted quarks: first results} 
 
\author{A. SHINDLER} 
 
\address{John von Neumann-Institut f\"ur Computing NIC, Platanenallee 6, D-15738 Zeuthen, Germany 
\\ E-mail: andrea.shindler@desy.de} 
 
\author{On behalf of the European Twisted Mass Collaboration (ETMC)}

\twocolumn[\maketitle\abstract{I report on first results of the ETMC obtained simulating  
  lattice QCD with two degenerate flavours of Wilson twisted mass fermions,
  for which physical observables are automatically O($a$) improved.  
  Recent improvements of the HMC algorithm allow simulations of  
  pseudoscalar masses below $300\, \mathrm{MeV}$ on volumes $L^3\cdot 2L$ with   
  $L> 2\, \mathrm{fm}$ and at  
  values of the lattice spacing $a\lesssim 0.1\, \mathrm{fm}$.}  
] 
\section{Lattice QCD} 
Simulations of lattice QCD are exact on the given lattice up to statistical errors, 
but many systematic uncertainties have to be controlled to have a reliable comparison between 
lattice QCD computations and experimental measurements. 
\vspace{-0.3cm} 
\begin{itemlist} 
 \item Renormalization and lattice artefacts: 
the choice of the lattice QCD action enters in the regularization of the theory. 
For this reason it is important to choose a lattice action that enjoys simple
renormalization properties and good scaling behaviour. 
 \item Light quark masses: it is crucial to use algorithms that allow simulations at light 
enough quark masses in order to eventually bridge 
the remaining gap between the simulation and the physical points, using chiral perturbation
theory\cite{Gasser:1983yg} ($\chi$PT). 
 \item Finite size effects: the volume has to be large enough in order to avoid systematic uncertainties 
stemming from the finite size of the system that is simulated.
\end{itemlist} 
The final goal is to simulate two  
light quarks and two non-degenerate heavier quarks keeping under control  
all these systematic uncertainties. 
Here I summarize first results for two light flavours obtained using a
theoretically well founded 
lattice QCD action and a highly improved algorithm. 
\section{Wilson twisted mass QCD} 
The lattice QCD action has a gauge $S_G$ and a fermionic $S_F$ part. 
The so called Wilson twisted mass (Wtm) action\cite{Frezzotti:2000nk} is 
\begin{equation} 
S_F = a^4\sum_x \big\{\bar\chi(x)[D_{\rm W}+i\mu\gamma_5\tau^3]\chi(x)\big\} ,
\end{equation} 
where  
\begin{equation} 
D_{\rm W} = \frac{1}{2} [\gamma_\mu(\nabla_{\mu} + \nabla_{\mu}^*) -  
a\nabla_{\mu}^* \nabla_{\mu}] + m_0 
\end{equation} 
is the Wilson-Dirac operator, 
$m_0$ is the untwisted (bare) quark mass, $\mu$ the (bare) twisted quark mass 
and $\tau_3$ the third Pauli matrix acting in flavour space. 
While parity-isospin is violated only at O($a^2$), the Wtm action has several
main advantages:
\begin{itemlist} 
  \item The twisted mass provides a sharp infrared cutoff for low eigenvalues\cite{Frezzotti:2000nk}. 
  \item Automatic O($a$) improvement when $m_0$ is properly tuned\cite{Frezzotti:2003ni}. 
  \item The renormalization pattern is in many cases greatly simplified\cite{Frezzotti:2004wz,Pena:2004gb}.
\end{itemlist} 
Note that also with the standard Wilson ($\mu=0$) formulation O($a$) improvement can be
obtained, but a whole set of parameters needs to be tuned to achieve this,
while with Wtm only one parameter needs to be tuned, $m_0$.
To be more specific automatic O($a$) improvement is at work if the value of $m_0$ is tuned to be the critical mass 
$m_{\rm c}$. In all the results presented here the critical mass is obtained
tuning the PCAC mass to zero and the role of the quark mass is played by the twisted mass $\mu$. 
For recent summaries on Wtm see the reviews~[\refcite{Frezzotti:2004pc,Shindler:2005vj}] and references therein. 
\vspace{-0.5cm} 
\subsection{Pseudoscalar decay constant}\label{subsec:fps} 
The pseudoscalar decay constant $f_{\rm PS}$ is an important quantity to compute  
for phenomenological applications (e.g. for the extraction of $V_{us}$ from the ratio $f_K/f_\pi$),  
and can be used to fix the lattice spacing in physical units. 
With Wtm this computation presents a set of advantages: 
no improvement coefficients are needed (namely $c_{\rm sw}$ and $c_{\rm A}$) due to automatic 
O($a$) improvement and moreover 
no renormalization factors ($Z_{\rm A}$) are needed\cite{Frezzotti:2000nk}.  
All the uncertainties related to the estimates  
of these parameters are simply absent. 
$f_{\rm PS}$ and many other physical quantities have been studied\cite{Jansen:2005kk} in the quenched model 
confirming that automatic O($a$) improvement is at work and that the O($a^2$) effects are small. 
\vspace{-0.5cm} 
\section{Algorithms and phase structure} 
To perform simulations at light quark masses with dynamical fermions,
a substantial amount of computer resources is required.
At the Lattice 2001 conference the situation was rather dramatic\cite{Ukawa:2002pc}. 
The low values of the quark masses, the large volume and the small lattice spacing 
reached in our work became possible only due to recent algorithmic developments\cite{Luscher:2005rx,Urbach:2005ji}. 
In particular, for all the $N_f=2$ results reported here 
we have used our variant of the HMC algorithm described in Ref.~[\refcite{Urbach:2005ji}]. 
In continuum QCD with $N_f > 1$ chiral symmetry is spontaneously broken at zero quark mass. 
The order parameter of this phase transition is the chiral condensate  
that is discontinous around the massless point. On the lattice Wilson-like actions break  
explicitly chiral symmetry even in the massless limit. It is then a natural question, if one is interested in 
simulations of light quark masses, to understand the chiral phase structure with Wilson-like quarks. 
Our collaboration has shown in a set of publications\cite{Farchioni:2004us,Farchioni:2005tu,Farchioni:2004fs} 
that there is a non-trivial phase structure at finite 
lattice spacing with a first order phase transition  
close to the chiral limit. These results are in agreement with one of the possible scenarios 
predicted\cite{Sharpe:1998xm,Munster:2004am} by lattice-$\chi$PT. 
In this scenario the first order phase transition line extends in the twisted
axis direction 
until a value of the twisted mass $\bar{\mu}$ of order $a^2$. As a consequence 
there is a minimal value of the quark mass that can be simulated at fixed
lattice spacing. Both the occurence of this scenario 
and the precise value of $\bar{\mu}$ depend on the details of the lattice
action, in particular on the choice of the gauge action $S_G$. 
The gauge actions so far studied by our collaboration can be parameterized by 
\vspace{-0.2cm} 
\begin{eqnarray} 
    S_{\rm G} &=& \beta \big[b_0\sum_{x;\mu<\nu}(1-{1 \over 3}P^{1 \times 
    1}(x;\mu,\nu)) +  \nonumber \\ 
&+& b_1 (1-{1 \over 3} P^{1 \times 2}(x;\mu,\nu)) \big] 
\end{eqnarray} 
with the normalization condition $b_0 = 1-8b_1$. $P^{1 \times 1}$ indicates the standard plaquette  
and $P^{1 \times 2}$ a planar $1 \times 2$ rectangular loop. 
In particular it has been shown\cite{Farchioni:2005ec} that the value of
$\bar{\mu}$ decreases if $b_1$ is properly chosen.
The choice of our collaboration was then to use $b_1=-1/12$ which corresponds to the so 
called tree-level Symanzik (tlSym) improved gauge
action\cite{Weisz:1982zw}. This choice compromises between a smaller value for
$\bar{\mu}$  than obtained with $b_1=0$ and avoiding problems with big scaling violation possibly  
induced by a too large value of $|b_1|$. 
\vspace{-0.5cm} 
\section{First results} 
\subsection{Setting the scale}\label{subsec:scale} 
In order to compare results obtained at different lattice spacings and  
to translate the simulations results in physical units, it is necessary to set the scale, i.e. 
to give the lattice spacing in physical units. Many choices are possible and some  
of them are under investigation by our collaboration. 
Here I present results obtained by using as an hadronic input the force between two static quarks 
at a certain intermediate distance $r_0$\cite{Sommer:1993ce}.  
While this quantity can be measured on the lattice very precisely, it has a rather uncertain 
phenomenological value. Nevertheless it can certainly be used to compare simulations at different lattice 
spacings. In this proceeding contribution to translate our results in physical units I use the value $r_0=0.5$ fm. 
\vspace{-0.5cm} 
\subsection{Pseudoscalar mass and decay constant}\label{subsec:obs} 
\begin{figure}[ht] 
\vspace{-0.5cm} 
\centerline{\psfig{file=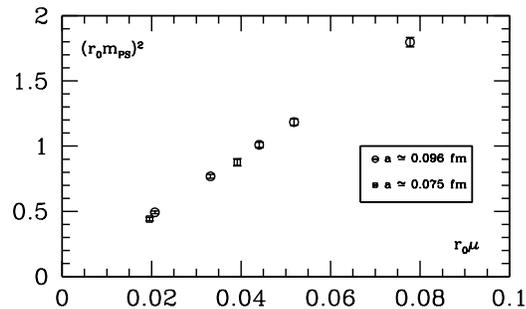,width=3.2in}} 
\vspace{-3.5cm} 
\caption{Quark mass dependence of the pseudoscalar mass squared.} 
\label{fig:mps} 
\vspace{-0.3cm} 
\end{figure}

\begin{figure}[ht] 
\centerline{\psfig{file=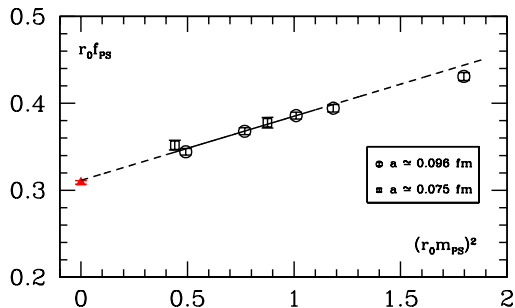,width=3.2in}} 
\vspace{-3.5cm} 
\caption{Dependence of the pseudoscalar decay constant on the squared  
pseudoscalar mass for 2 different lattice spacings.  
As a comparison is also shown the phenomenological value (${\color{red}\blacktriangle}$)  
in the chiral limit, and a linear fit to the 5 lightest  
pseudoscalar masses.} 
\label{fig:fps} 
\end{figure}

\begin{table}[t] 
  \centering 
\tbl{The parameters for the simulation.  
    In order to convert to physical units,  
    I take the value of $r_0/a$ as computed at the minimal values  
of the twisted mass simulated, that can be read from Table~\ref{tab:results}. 
  \label{tab:setup}} 
{  \begin{tabular*}{1.\linewidth}{@{\extracolsep{\fill}}lccc} 
    \hline\hline 
    $\Bigl.\Bigr.\beta$ & $L^3\times T$  
    & $r_0/a(a\mu_\mathrm{min})$ & $a$ [fm] \\ 
    \hline\hline 
    $3.9$  & $24^3\times48$ & $5.184(41)$   & $\simeq 0.096$ \\ 
    $4.05$ & $32^3\times64$ & $6.525(101)$  & $\simeq 0.075$ \\ 
    \hline\hline 
  \end{tabular*}} 
\vspace{-0.3cm} 
\end{table} 
The current status of our simulations are summarised in Tables 
\ref{tab:setup} and \ref{tab:results}.  
In Table \ref{tab:setup} the parameters of our simulations are provided, 
and in Table \ref{tab:results} 
current estimates of the pseudoscalar mass $m_\mathrm{PS}$ and the
renormalized quark mass in physical units are given. A rough estimate of the missing
renormalization factor can be obtained from our data, using as inputs the
available lattice estimates for the strange quark mass and $m_s/m_{u,d} \simeq 26$.
\begin{table}[t] 
  \centering 
\tbl{Current status of the simulations at $\beta=3.9$ and $\beta=4.05$ using 
    the parameters in 
    Table~\ref{tab:setup}: current estimates 
    of the pseudoscalar and the renormalized quark masses, 
    the number of trajectories generated after $1500$ trajectories of
    thermalization.
  \label{tab:results}}
{  \begin{tabular*}{1.\linewidth}{@{\extracolsep{\fill}}lccc} 
    \hline\hline 
    $\Bigl.\Bigr.\beta$ & $\mu_{\rm R}$ [MeV] 
    & $m_\mathrm{PS}\,[\mathrm{MeV}]$ &  
    $N_\mathrm{traj}$ \\ 
    \hline\hline 
    $3.9$ & $\simeq 18$ & $\simeq 270$ & $5000$ \\ 
 
    & $\simeq 29$ & $\simeq 350$  & $5000$ \\ 
 
    & $\simeq 38$ & $\simeq 400$  & $5000$ \\ 
 
    & $\simeq 45$ & $\simeq 430$  & $5000$ \\ 
 
    & $\simeq 67$ & $\simeq 530$  & $5000$ \\ 
    \hline 
    $4.05$ & $\simeq 18$ & $\simeq 270$  & $4200$ \\ 
 
    & $\simeq 34$ & $\simeq 370$ & $3000$ \\ 
    \hline\hline 
  \end{tabular*}} 
\vspace{-0.3cm} 
\end{table} 
In Figure~\ref{fig:mps} the quark mass dependence of the pseudoscalar mass
squared is shown. 
The dependence is to a good approximation linear, and the results at different  
lattice spacings do not show appreciable cutoff effects. I recall that the twisted quark mass 
in this plot still needs to be renormalized. 
In Figure~\ref{fig:fps} I show the dependence of the pseudoscalar decay constant on the squared  
pseudoscalar mass. 
The rather consistent values obtained 
at the two different lattice spacings is a very promising feature as far as
the control on the systematic uncertainties of this physical quantity is concerned.  
For both the plots I have used $\chi$PT in order to correct for finite size effects\cite{Colangelo:2005gd}.  
In a forthcoming publication\cite{ETMC} we will analyze these data according to $\chi$PT. Here in Fig.~\ref{fig:fps} 
I simply show for illustration a linear fit to the 5 lightest  
pseudoscalar masses, and the phenomenological estimate\cite{Gasser:1983yg,Colangelo:2003hf}  
of the pseudoscalar decay constant in the chiral limit. 
\vspace{-0.7cm} 
\section{Conclusions and outlooks} 
The control of systematic uncertainties on lattice computations is a prerequisite in order to compare them 
with experimental measurements. I have summarized here some properties of Wtm,  
together with some first results with $N_f=2$ light dynamical quarks.  
These results (see also the proceedings\cite{Jansen:2006rf} at the Lattice 2006 conference)  
are rather encouraging and suggest that the systematic uncertainties can be
controlled simulating improved lattice actions,  
with the newly developed algorithms on currently available supercomputers.

\vspace{-0.5cm} 
\section*{Acknowledgments} 
\vspace{-0.1cm} 
It is a pleasure to aknowledge all the members of the ETM Collaboration that 
have contributed to the results presented here. 
I thank NIC and the computer centres at Forschungszentrum J{\"u}lich 
and Zeuthen for providing the necessary technical help and computer 
resources. Computer time on UKQCD's QCDOC in Edinburgh\cite{Boyle:2005fb}  
using the Chroma code\cite{Edwards:2004sx} and on 
apeNEXT\cite{Bodin:2005gg} in Rome and Zeuthen and on MareNostrum in 
Barcelona (www.bsc.es) are gratefully acknowledged. 
This work has been supported in part by the DFG  
Sonderforschungsbereich/Transregio SFB/TR9-03  
and the EU Integrated Infrastructure Initiative Hadron Physics (I3HP) 
under contract RII3-CT-2004-506078.  
I also thank the DEISA Consortium (co-funded by the EU, FP6 project 508830), 
for support within the DEISA Extreme Computing Initiative (www.deisa.org).  
 
\balance 
 
\bibliographystyle{JHEP} 
\bibliography{proc}

\end{document}